%% file: main.tex
\documentclass[letterpaper, 10 pt, conference]{ieeeconf}  

\IEEEoverridecommandlockouts                             

\usepackage{graphics}
\usepackage{epsfig}
\usepackage{times}
\usepackage{amsmath}
\usepackage{xcolor}

\usepackage{mathrsfs}
\usepackage{amssymb}
\usepackage{cite}
\usepackage{dsfont}
\usepackage{xcolor}
\usepackage{mathtools}
\usepackage{mathrsfs}
\usepackage{subfig}
\usepackage{balance}
\usepackage[ruled]{algorithm2e}
\usepackage{tcolorbox}
\usepackage{bbm}
\newtheorem{remark}{Remark}

\usepackage{xurl}

\title{\LARGE \bf
Reachability Analysis for Black-Box Dynamical Systems
}
\author{Vamsi Krishna Chilakamarri$^*$$^{1}$, Zeyuan Feng$^*$$^{2}$, and Somil Bansal$^{2}$
\thanks{$^*$ Denotes equal contribution. $^{1}$Author is with Indian Institute of Technology, Madras, India. $^{2}$Authors are with ECE at the University of Southern California, LA, USA. Corresponding author: \{zeyuanfe@usc.edu\}.} 
\thanks{This work is supported by NSF CAREER and DARPA ANSR Programs.}
\thanks{ Project website: \protect\url{https://github.com/sia-lab-git/blackbox_reachability}}
}

\begin{document}

\maketitle
\thispagestyle{empty}
\pagestyle{empty}

\begin{abstract}
Hamilton-Jacobi (HJ) reachability analysis is a powerful framework for ensuring safety and performance in autonomous systems. 
However, existing methods typically rely on a white-box dynamics model of the system, limiting their applicability in many practical robotics scenarios where only a black-box model of the system is available. 
In this work, we propose a novel reachability method to compute reachable sets and safe controllers for black-box dynamical systems. 
Our approach efficiently approximates the Hamiltonian function using samples from the black-box dynamics. 
This Hamiltonian is then used to solve the HJ Partial Differential Equation (PDE), providing the reachable set of the system. 
The proposed method can be applied to general nonlinear systems and can be seamlessly integrated with existing reachability toolboxes for white-box systems to extend their use to black-box systems.  
Through simulation studies on a black-box slip-wheel car and a quadruped robot, we demonstrate the effectiveness of our approach in accurately obtaining the reachable sets for black-box dynamical systems.
\end{abstract}

\section{Introduction}
\label{sec:intro}
\input{intro}

\section{Problem Setup}
\label{Problem Setup}
\input{problem}

\section{Background: Hamilton-Jacobi Reachability}
\label{sec:background}
\input{background}

\section{Solving HJB-VI for Black-Box Dynamics}
\label{sec:approach}
\input{approach}

\section{Experiments}
\label{sec:results}
\input{results}

\section{Discussion and Future Work}
\label{sec:conclusion}
\input{conclusion}

\bibliographystyle{IEEEtran}
\bibliography{references}

\end{document}

%% file: intro.tex
As robots operate in increasingly complex environments, ensuring safe interactions with their surroundings is crucial. 
A widely used approach to design safe controllers for robotic systems is reachability analysis \cite{8263977}, which involves computing the Backward Reachable Tube (BRT) of the system. 
Intuitively, the BRT represents the set of all initial states from which the system will inevitably enter a failure set, such as obstacles for a navigation robot, despite the best control effort. 
Therefore, the complement of the BRT defines the safe states for the system.

Several methods have been developed to compute BRTs for dynamical systems, including approaches that approximate BRTs using Zonotopes \cite{althoff2010computing, schurmann2017guaranteeing}, Sum-of-Squares programming \cite{Majumdar13, majumdar2017funnel}, and parallelotopes \cite{Dreossi16}.
For a comprehensive survey on BRT computation methods, we refer interested readers to \cite{9561949, 8263977}.
Methods that compute BRTs accurately up to numerical precision include level set methods \cite{mitchell2005time,lygeros2004reachability,fisac2015reach,liao2024improved}.
Level set methods are rooted in Hamilton-Jacobi (HJ) Reachability analysis, where BRT computation is formulated as an optimal control problem. 
This amounts to solving a partial differential equation, called the HJ-PDE.  
Correspondingly, techniques have been developed to solve this PDE numerically over a grid \cite{mitchell2004toolbox, pythonhjtoolbox} or through learning-based methods \cite{9561949, darbon2020overcoming, niarchos2006neural}.

However, most of these methods rely on an analytical (white-box) dynamics model of the system to compute the BRT. 
Unfortunately, as robotic systems grow increasingly complex, this requirement becomes prohibitive. 
For instance, recent advances in simulation technologies \cite{howell2022predictive, makoviychuk2021isaac} allow for the simulation of complex dynamical systems, but the underlying dynamics are often available only as black-box models. 
This limitation renders current reachability methods ineffective for the safety analysis of such systems.

Several methods have been developed to compute BRTs and design safe controllers for black-box dynamical systems.
These methods can be broadly classified into model-based and model-free approaches.
Model-based methods first obtain or learn a white-box model of the system, possibly with uncertainty bounds, and then apply traditional reachability techniques \cite{fliess2006complex, herbert2021scalable,fisac2018general,akametalu2014reachability, wang2024providing}. 
While promising, the accuracy of the computed BRT heavily depends on the accuracy of the learned dynamics model, which can undermine the reliability of the resultant safety assurances.
Other approaches leverage side information about the black-box dynamics to compute reachable sets, such as assuming linear time-invariant dynamics \cite{pmlr-v134-chen21c}, utilizing Lipschitz constants of the dynamics and state monotonicity \cite{djeumou2021fly,djeumou2021fly2}, or employing polynomial templates to construct reachable sets \cite{kapinski2014simulation,topcu2008local,ravanbakhsh2019learning}.

\begin{figure*}
    \centering
    \includegraphics[width=0.8\linewidth]{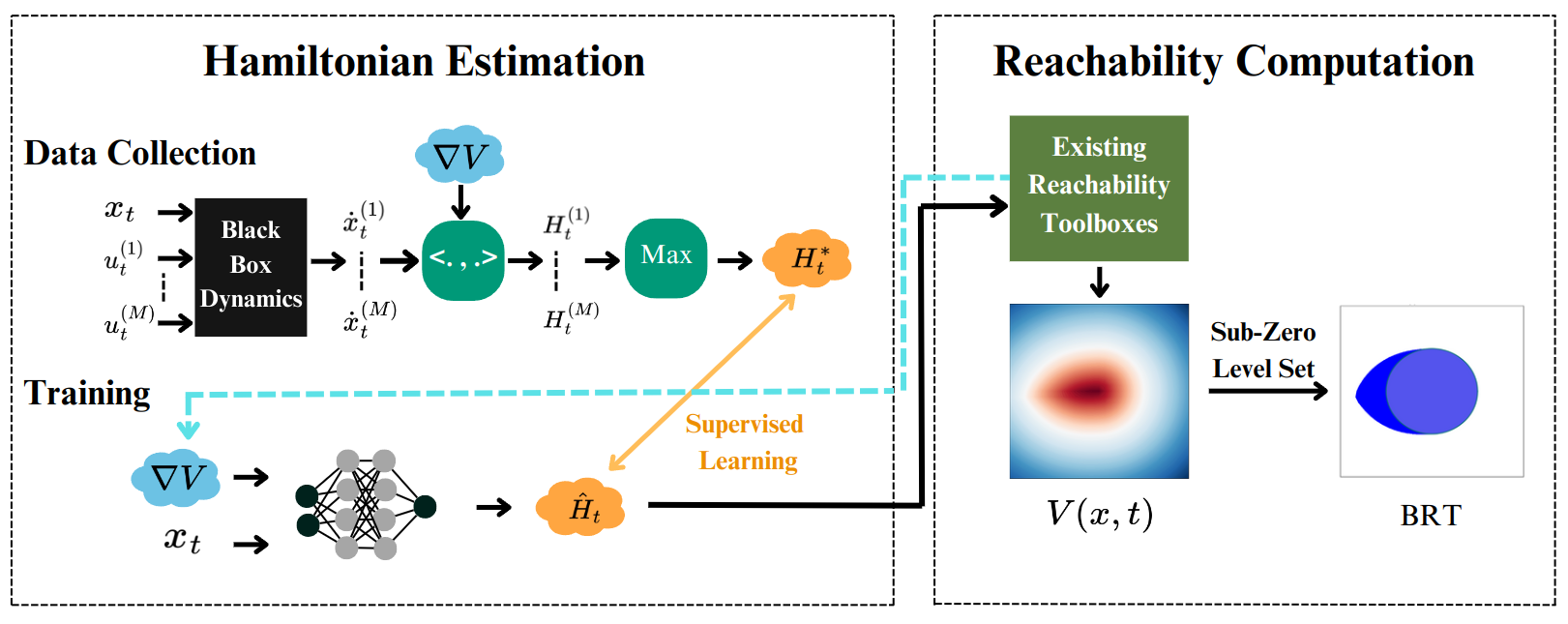}
    \caption{An overview of the proposed method: we use the samples from the black-box dynamics to construct a dataset where each sample consists of state $x_t$, the spatial derivative of value function $\nabla V$, and the corresponding Hamiltonian label $H^*$. This dataset is then used to train a neural network to estimate the Hamiltonian at any given state and $\nabla V$. The trained network can be integrated within existing reachability toolboxes to solve the Hamilton-Jacobi PDE to compute the value function and the BRT for black-box dynamical systems.}
    \vspace{-1em}
    \label{fig:overview}
\end{figure*}

In contrast, model-free methods do not explicitly construct a white-box model of the system.
These include sampling-based techniques \cite{liebenwein2018sampling}, such as scenario optimization \cite{sutter2017data, devonport2020estimating, ghosh2019new} and adversarial sampling methods \cite{lew2020sampling}.
Another line of model-free approaches uses Reinforcement Learning (RL) to estimate HJ reachability solutions, thereby naturally handling unknown system dynamics \cite{fisac2019bridging, hsu2021safety, hsu2023isaacs}. 
While promising for black-box reachability problems, RL methods typically lack safety assurances, especially when function approximations are used, and the accuracy of the BRT is sensitive to the choice of discount factor.
Additionally, these methods are not compatible with existing reachability techniques, such as level-set toolboxes \cite{mitchell2004toolbox, pythonhjtoolbox}.

In this work, we extend level set methods to compute reachable sets for black-box dynamics models. 
Our \textit{key idea} is to construct a zeroth-order approximation of the Hamiltonian function in the HJ-PDE using samples from the black-box dynamics.
This approximation, which can be efficiently computed with modern computational tools, is then used to solve the HJ-PDE and approximate the BRT for the black-box system.
We also provide an algorithm to further improve the efficiency of the approximation if the underlying black-box system is control-affine. 

Our method can be readily integrated with existing reachability toolboxes -- such as grid-based level-set toolboxes \cite{mitchell2004toolbox, pythonhjtoolbox} and learning-based methods \cite{9561949} -- to synthesize BRTs and optimal safe policies for black-box dynamical systems.
An overview of the proposed method is illustrated in Fig.~\ref{fig:overview}.
We demonstrate the implementation of our approach in multiple reachability computation tools, including a bicycle robot system using the level-set toolbox \cite{pythonhjtoolbox}, and a slip-wheel car system and a high-dimensional quadruped system using DeepReach \cite{9561949}, a learning-based reachability framework.
In all cases, we compare our approach with various model-based and model-free methods and show its effectiveness in providing a more accurate approximation of the BRT.

%% file: problem.tex
Consider a black-box dynamical system with \textit{unknown}, possibly nonlinear, time-invariant dynamics: $\dot{x}=f(x,u)$,
where $x \in \mathbb{R}^n$ is the state, and $u \in \mathcal{U} \subset \mathbb{R}^{n_u}$ is the control input.
Although $f$ is unknown, we assume that we can query the system to obtain the state transition $(x(t),u(t),x(t+\Delta))$ (e.g., through a simulator), where $\Delta$ is the time step. 
We are also given a failure set $\mathcal{L} \subset \mathbb{R}^n$ that represents undesirable states for the system (e.g., obstacles for a mobile robot), and thus, needs to be avoided. 

Our main goal in this work is to compute the backward reachable tube (BRT) of the system, $\mathcal{B}$, defined as the set of all initial states from which the system will inevitably enter $\mathcal{L}$ within the time horizon $[0, T]$, regardless of the control strategy.
Mathematically,
\begin{equation}
\label{eq:avoid_BRT}
    \mathcal{B}=\left\{x: \forall u(\cdot) \in \mathcal{U}, \exists \tau \in[0, T], \xi_{x,0}^{u(\cdot)}(\tau) \in \mathcal{L} \right\},
\end{equation}
where $\xi_{x,t}^{u(\cdot)}(\tau)$ denote the system state achieved at time $\tau$ by applying the policy $u(\cdot)$, starting from an initial state $x$ at time $t$.
As such, the BRT contains the states that are unsafe for the system and should be avoided.
Correspondingly, our second goal is to synthesize a safe control policy $u^*(\cdot)$ for the system that prevents the system from entering $\mathcal{B}$.

\vspace{0.2em}

\noindent \textbf{\textit{Running Example: A Bicycle Robot.}}
We now introduce a running example to explain the key ideas.
The robot is modeled as a 5-dimensional system with state
$x = [p_x, p_y, v, \phi, \delta]$, where $p_x$, $p_y$ is the planar position, $v$ is velocity, $\phi$ is the heading, and $\delta$ is the steering angle. 
The robot control is given by $u = [a, \omega]$, where $a \in [-2, 2] \, \text{m/s}^2$ is the linear acceleration and $\omega \in [-2, 2] \, \text{rad/s}$ is the angular velocity. We use $L = 1$ in our case study. 
The robot dynamics are control affine and are modeled as:
\begin{equation*}
\dot{p}_x = v \cos \phi,~\dot{p}_y = v \sin \phi,~\dot{v} = a,\\ 
~\dot{\phi} = \frac{v}{L} \tan \delta,~\dot{\delta} = \omega
\end{equation*}
However, to test our methods, we assume that the dynamics are not known explicitly and are presented as a black box to us. Specifically, given a state $x$ and control input $u$, the black-box dynamics simulates the system for a timestep $\Delta = 0.001 \text{s}$, and outputs the resultant state $x(t+\Delta)$.
The failure set $\mathcal{L}$ is given by a circle of radius 2.5$\text{m}$ centered at the origin. 
Our goal is to compute the BRT of the system corresponding to this failure set while only relying on the black-box dynamics.

%% file: background.tex
HJ Reachability analysis is a powerful tool for synthesizing BRTs for dynamical systems \cite{mitchell2005time}. 
Typically, users start with specifying a target function $l: \mathbb{R}^n \rightarrow \mathbb{R}$, with its sub-zero level set being the failure set, i.e., $\mathcal{L} = \{x:  l(x) \leq 0\}.$
Given $l$, the BRT computation is formulated as an optimal control problem with the cost function: 
\begin{equation}
    J(x, t, u(\cdot)) =  \min _{\tau \in[t, T]} l\left(\xi_{x,t}^{u(\cdot)}(\tau)\right).
\end{equation}
Intuitively, $J \leq 0$ implies that the system entered the failure set during the time horizon $[t, T]$ under $u(\cdot)$.
Thus, to capture safety violations under an optimal policy, we can compute the corresponding value function:
\begin{equation}
    V(x, t) =  \sup _{u(\cdot) \in \mathcal{U}_{[t, T]}} \min _{\tau \in[t, T]} l\left(\xi_{x,t}^{u(\cdot)}(\tau)\right).
\end{equation}
The value function can be obtained by solving the Hamilton-Jacobi-Bellman Variational Inequality (HJB-VI):
\begin{equation}
\label{eq:HJB-VI}
\begin{gathered}
     \min \{D_{t}V(x,t) + H(x,t), l(x) - V(x,t) \} = 0, \\
     V(x,T) = l(x),
\end{gathered}
\end{equation}
where $H(x,t)$ is called the Hamiltonian and encodes the role of system dynamics:
\begin{equation}
\label{eq:Ham}
    H(x,t) = \max_{u \in \mathcal{U}} \langle \nabla V(x,t) \; , \; f(x,u) \rangle.
\end{equation}
Here, $D_{t}$ and $\nabla$ denote the time and spatial derivatives of the value function.
We refer the interested readers to \cite{mitchell2004toolbox, bansal2017hamilton} for a more detailed exposition of reachability analysis.

Once the value function is computed, the BRT is given by the sub-zero level set of the value function:
\begin{equation}
\label{eq:brt_from_V}
\mathcal{B}=\left\{x: V(x,0)  \leq 0 \right\}.
\end{equation}
Along with the BRT, the value function provides an optimal safety controller to keep the system outside the BRT:
\begin{equation}
    u^*(x,t)=\arg \max _u \langle\nabla V(x, t), f(x, u)\rangle.
\end{equation}

\subsection{An Overview of Methods to Solve the HJB-VI}
Traditionally, numerical methods are employed to solve the HJB-VI over a grid representation of the state space \cite{mitchell2004toolbox, pythonhjtoolbox}, wherein the time and spatial derivatives in \eqref{eq:HJB-VI} are approximated numerically over the grid.
These methods rely on an analytical expression of $f$ to compute the Hamiltonian, i.e., to solve the optimization problem in \eqref{eq:Ham}.
This is the source of the key challenge in applying these methods to black-box dynamical systems.

While the grid-based methods offer accurate solutions for low-dimensional problems, they suffer from the curse of dimensionality. Consequently, learning-based methods have been developed to solve HJB-VI for high-dimensional cases. 
Here, we present an overview of one such method, DeepReach \cite{9561949}, which we will use in some of our case studies.
DeepReach uses a self-supervised learning scheme to estimate the solution of HJB-VI.
In particular, the value function is approximated as 
$V_\theta(x,t)=l(x)+(T-t)\cdot O_\theta(x,t)$, where $O_\theta(x,t)$ is the output of a sinusoidal neural network (NN) and $\theta$ represents trainable parameters.
The loss function to train the NN is given by the HJB-VI residual error:
\begin{equation} \label{deepreach_loss}
\begin{aligned}
& h_{1}\left(x_i, t_i ; \theta\right)=\| \min \left\{D_t V_\theta\left(x_i, t_i\right)+H\left(x_i, t_i\right),\right. \\
& \qquad \qquad \left.l\left(x_i\right)-V_\theta\left(x_i, t_i\right)\right\} \|
\end{aligned}
\end{equation}
Once again, the loss function in \eqref{deepreach_loss} requires computing the Hamiltonian, which in turn, relies on a white-box dynamics model. 
In this work, we aim to alleviate this limitation.

%% file: approach.tex
This section presents an approach to compute BRTs for general black-box dynamical systems and then discuss a tailored approach under a control-affine assumption. 

\subsection{A General Approach to Compute BRTs for Black-Box Dynamical Systems}\label{section: BRT_computation}
Our approach consists of computing an approximation of the Hamiltonian function using samples of system dynamics. The approximated Hamiltonian is then used in the HJB-VI in \eqref{eq:HJB-VI} to compute the value function and corresponding BRT.

\vspace{0.5em}
\noindent \textbf{\textit{Hamiltonian Estimation.}} We propose to train a NN, denoted as $H_\nu$, to predict the Hamiltonian given a state $x$ and the spatial gradient of the value function, $\nabla V(x,t)$.
Using a NN allows for a quick inference of the Hamiltonian at the grid points (for numerical methods) or at the data samples (for learning-based methods) during the value function computation, though other function approximations can also be used.

To train the Hamiltonian estimator, we collect a dataset $\mathcal{D}_H$ as follows: we randomly sample a batch of (normalized) spatial gradient vectors and system states, $\left\{[x \ \nabla \bar{V}(x,t)]\right\}$.
Next, for each state-gradient pair, we randomly sample a set of control inputs and query the corresponding next states of the system through black-box model in a parallel manner. 
Finally, we compute the Hamiltonian corresponding to each control sample using \eqref{eq:Ham} and compute the empirical maximum, $\bar{H}$.  
The detailed dataset collection procedure is described in Algorithm \ref{alg:dataset}.
Once the dataset is collected, the Hamiltonian estimator is trained via supervised learning to optimize: 
\begin{equation}
\label{hamloss}
h_{H}\left(x_i, \nabla \bar{V}(x,t)_i ; \nu\right) :=\| H_\nu(x_i, \nabla \bar{V}(x,t)_i) -\bar{H} \|.
\end{equation}
After being trained, the parameters $\nu$ are frozen when $H_\nu$ is used to predict the Hamiltonian during the value function computation.
The Hamiltonian prediction is given by
\begin{equation}
\hat{H} \left( x, \nabla V(x,t); \nu \right) =\|\nabla V(x,t)\| H_\nu \left(x, \frac{\nabla V(x,t)}{\|\nabla V(x,t)\|} \right),
\end{equation}
where we account for the normalization of $\nabla V(x,t)$.
We next solve the HJB-VI in \eqref{eq:HJB-VI} as usual with $H$ replaced by $\hat{H}$ to obtain an approximation $\hat{V}(x, t)$ of the value function.
\begin{remark} \label{remark:1}
Note that the actual distribution of $\nabla V(x,t)$ during the value function calculation can be different from the uniform distribution that is used to collect the data to train $H_\nu$.
This can cause inaccuracies in the predicted Hamiltonian, especially when the underlying dynamics are high-dimensional or stiff. 
For these reasons, we compute the value function once and augment the dataset $\mathcal{D}_H$ with data points $(\left[x \ \nabla \bar{V}(x,t)\right],\bar{H})$, where the spatial derivative is sampled around the computed $\nabla \hat{V}(x,t)$.
With the augmented dataset, we repeat the procedure to train $H_\nu$ and recompute the value function.
\end{remark}

\vspace{0.5em}
\noindent \textbf{\textit{Obtaining the Optimal Safe Policy.}}
The optimal safety control is obtained by training a neural network controller $u_\psi$ using the dataset $\mathcal{D}_H$. The network takes $x$ and $\nabla V(x,t)$ as inputs and predicts the optimal safety control. 
To train the controller network, we optimize the Mean Absolute Error loss between the normalized optimal control labels $\hat{u}^*(x)$ and the predicted control
\begin{equation}
     h_c(x_i,t_i; \psi) =  \left\lVert u_\psi(x_i,\nabla V(x_i,t_i))-\hat{u}^*(x_i)   \right\rVert.
\end{equation}

\begin{algorithm}[h!]
\caption{Data Collection for Ham Estimation}\label{alg:dataset}
\textbf{Input:} A pre-collected set of states $\mathcal{D}_X$, e.g., states sampled uniformly over a state space\;
\textbf{Output:} $\mathcal{D}_H$ \;
\textbf{Parameters:} simulator time step $\Delta$, number of samples, number of control samples\;
Initialize $\mathcal{D}_H \leftarrow \emptyset$\;
\ForEach{sample}{
    Sample $x \overset{\mathrm{iid}}{\sim} \text{Uniform}(D_X)$\;
    Sample $\nabla V(x,t) \overset{\mathrm{iid}}{\sim} \text{Uniform}(\mathcal{C})$, $ \mathcal{C}=\left\{ x \in \mathbb{R}^n \mid |x_i| \leq 1, \forall i=1,2,...,n \right\} $\;
    $\nabla \bar{V}(x,t) \leftarrow \frac{\nabla V(x,t)}{\| \nabla V(x,t)\|} $ \;
    $\bar{H} \leftarrow -\infty$ \;
    \ForEach{control sample}{
        $u \overset{\mathrm{iid}}{\sim} \text{Uniform}(\mathcal{U})$\;
        Take action $u$ from $x$, observe $x_{next}$\;
        \uIf{$\bar{H} < \langle \nabla V(x,t) \; , \; \frac{x_{next}-x}{\Delta} \rangle$}{
            $\bar{H} \leftarrow \langle \nabla V(x,t) \; , \; \frac{x_{next}-x}{\Delta} \rangle$, \quad $u^* \leftarrow u$ \;
          }
    }
    $\mathcal{D}_H \leftarrow \mathcal{D}_H  \cup \left\{\left(\left[x \ \nabla \bar{V}(x,t)\right],\bar{H}, u^*\right)\right\}$ 
}
\end{algorithm}

\noindent \textbf{\textit{Safety Assurances for the Obtained BRT.}} 
To reason about safety assurances under the proposed framework, we leverage the formal verification method proposed in \cite{lin2024verification} to obtain a probabilistic safe set from an approximate value function and corresponding optimal policy, $(\hat{V},u_\psi)$.
The overall idea is to provide a high-confidence bound $\delta$ on the value function error using conformal prediction. 
This results in a correction of the value function by $\delta$.
The corrected value function is then used to compute the BRT and the safe set.

Specifically, the method requires users to specify a confidence parameter $\beta \in (0,1)$ and a safety violation parameter $\epsilon \in (0,1)$. It then computes an error bound $\delta$ using conformal prediction to ensure that with at least $1-\beta$ probability: 
\begin{equation}
    \underset{x \in \mathcal{S}}{\mathbb{P}}( \min _{\tau \in[0, T]} l\left(\xi_{x, 0}^{u_\psi}(\tau)\right) \leq 0) \leq \epsilon,
\end{equation}
where $\mathcal{S}=\left\{x: x \in \mathcal{X}, \hat{V}(x,0)>\delta \right\}$. In other words, the probability of a state within the super-$\delta$ level set of $\hat{V}$ being actually unsafe during the rollouts under $u_\psi$ is at most $\epsilon$. 
The complement of $\mathcal{S}$, denoted as $\mathcal{B}_\epsilon$, thus represents a high-confidence estimate of the BRT. 

\subsection{Solving HJB-VI for Control-Affine Black-Box Systems} \label{sec:control_affine_variant}
We now propose a variant of our approach to compute the value function under the assumption that the underlying black-box system is control-affine with box-constrained control inputs.
Specifically, we now consider systems with dynamics $f(x, u) := f_1(x) + f_2(x)u$, with the control input $u := [u_1, \ldots, u_{n_u}] \in [\alpha_i- \beta_i, \alpha_i+ \beta_i]^{n_u}$.
This formalism aligns well with many real-world robotic systems, which are often control-affine in nature. 

Our key observation is that for control-affine systems, the Hamiltonian in \eqref{eq:Ham} is optimized at one of the extremal control inputs \cite{mitchell2005time}.
This is because the objective function in \eqref{eq:Ham} is linear with respect to control variables $u_1, u_2, \ldots, u_{n_u}$ for control-affine systems, subjected to an $n_u$-dimensional hypercube centered at $(\alpha_1, \alpha_2, \ldots, \alpha_{n_u})$ and side length $\beta_i$ in the $i$th dimension.
Consequently, the optimization of the Hamiltonian reduces to a linear program (LP), whose solution lies at one of the corners of the hypercube.
Thus, the Hamiltonian can be computed as:
\begin{equation} \label{ham:control_affine}
    H(x, t) = \max_{u \in \{\alpha_i - \beta_i, \alpha_i + \beta_i\}^{n_u}} \langle \nabla V(x,t) \; , \; f(x, u) \rangle
\end{equation}
The above equation suggests a tractable mechanism for evaluating the Hamiltonian for black-box control-affine systems.
Specifically, by restricting the evaluation to a finite set of control inputs corresponding to the vertices of the hypercube, the Hamiltonian can be accurately obtained by using the control samples at these vertices.
Thus, the Hamiltonian can be estimated as:
\begin{equation} \label{ham:control_affine_estimation}
\hat{H} \left( x, \nabla V(x,t)\right) = \max_{u \in \{\alpha_i - \beta_i, \alpha_i + \beta_i\}^{n_u}} \langle \nabla V(x,t)\; , \frac{x_{next}^u-x}{\Delta}\; \rangle
\end{equation}
where $x_{next}^u$ is the next state under control input $u$. 
Unlike the method proposed in Sec. \ref{section: BRT_computation}, the proposed variant bypasses any need for data collection or network training for estimating the Hamiltonian, thereby avoiding any value function errors emanating from incorrect Hamiltonian estimation for control-affine systems. 

\begin{remark}
    If $\Delta$ is sufficiently small, the Hamiltonian estimation error tends to zero, and the BRT computed using the proposed scheme converges to the ground truth BRT. This follows immediately since as $\Delta \rightarrow 0$, the dynamics flow estimation becomes more accurate.
\end{remark}

\begin{remark}
    Even when the black-box system is not control-affine, the proposed variant generates a provably conservative BRT for the system because it always underapproximates the true Hamiltonian.
\end{remark}

\begin{remark}
    A key limitation of the proposed variant is that it is not scalable to high-dimensional input spaces, since it requires sampling $2^{n_u}$ controls in each iteration. 
    To address this issue, we can individually determine the value of optimal control in each input dimension.
    Specifically, we sample a random nominal control input $\boldsymbol{u}_{nom}=(u_1^{nom},u_2^{nom},...,u_{n_u}^{nom})$, and $n_u$ extra controls given by $\boldsymbol{u}_{i}=(u_1^{nom},...,\alpha_i + \beta_i,...,u_{n_u}^{nom})$, respectively.
    The optimal control in the $i^{th}$ dimension can be obtained by comparing the Hamiltonian at the nominal control input and $\boldsymbol{u}_{i}$:
    \begin{equation}
        u^*_{i}(x)= 
        \begin{cases}
            \alpha_i + \beta_i,& \text{if } 
            \langle \nabla V(x,t)\; , \frac{x_{next}^{\boldsymbol{u}_{i}}-x_{next}^{\boldsymbol{u}_{nom}}}{\Delta}\; \rangle \geq 0\\
            \alpha_i - \beta_i,              & \text{otherwise}
        \end{cases}
    \end{equation}
    The above method reduces the number of control samples from $2^{n_u}$ to $(n_u+1)$, thereby extending the scalability to high-dimensional control spaces. 
\end{remark}

\noindent \textbf{\textit{Obtaining Safe Policy.}}
To obtain the safe policy, we can follow the same procedure as outlined in Sec. \ref{section: BRT_computation}. However, since the optimal control is extremal in this case, one can also use a classifier network that predicts one of the two classes $\{\alpha_i + \beta_i, \alpha_i - \beta_i\}$ for the $i^{th}$ dimension.

\subsection{Running Example}
For the general approach (\textbf{\textit{referred to as Ham-NN here on}} for \underline{Ham}iltonian via \underline{N}eural \underline{N}etwork), we approximate $H_\nu$ with a NN with 2 hidden layers and 64 neurons in each layer with ReLU activation.
For training, we collected a dataset of 500K samples.
For each state-gradient datapoint, we sample 10K controls to find the optimal control and Hamiltonian for training.
Once collected, the NN was trained on this dataset using the loss function \eqref{hamloss}. 
Dataset collection took $\approx 5$ mins, and the network training too $20$ mins on a single NVIDIA GeForce RTX 3090 GPU worker.
Once trained, the predicted Hamiltonian from the network was used to solve the HJB-VI and compute the value function using the Python Level Set Toolbox \cite{pythonhjtoolbox} over a grid of size [31, 31, 21, 51, 11].

Since the dynamics are control-affine in this case, we also compute the value function using the variant proposed in Sec. \ref{sec:control_affine_variant}.
For this variant (\textbf{\textit{referred to as Ham-CA here on}} for \underline{Ham}iltonian via \underline{C}ontrol \underline{A}ffine dynamics), we sample the four corners of the 2D control hypercube and query the next state to estimate the Hamiltonian as per 
\eqref{ham:control_affine_estimation}. 
Finally, for comparison purposes, we also compute the ground truth BRT using the actual dynamics.
We evaluate the Mean Squared Error (MSE) of the Value function and the False Positive rate (FP$\%$) for both approaches against the ground truth.

\begin{figure}[!t]
    \centering
    \subfloat[Ground Truth]{\includegraphics[width=0.30\linewidth]{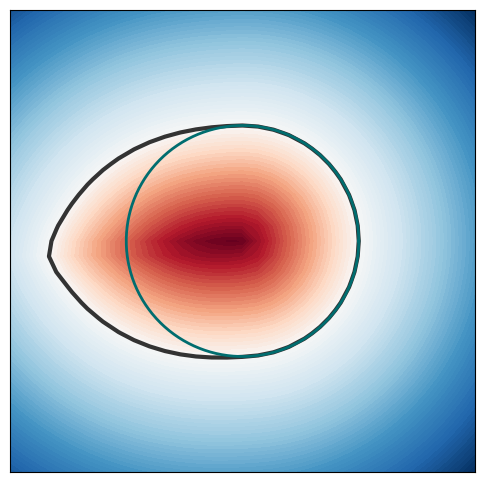}\label{fig:ground_truth}}
    \hfill
    \subfloat[Ham-NN]{\includegraphics[width=0.30\linewidth]{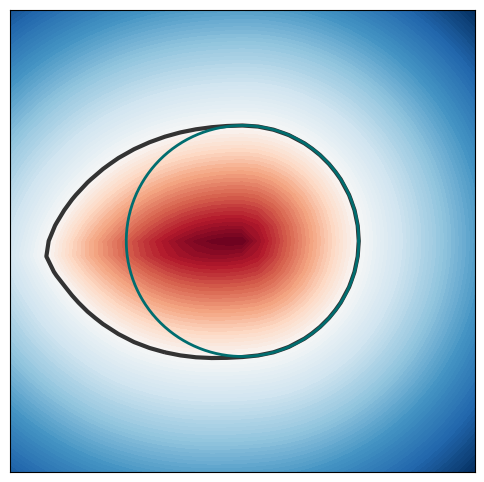}\label{fig:general_approach}}
    \hfill
    \subfloat[Ham-CA]{\includegraphics[width=00.30\linewidth]{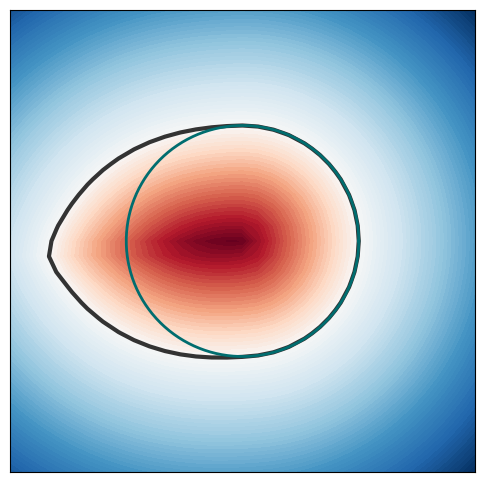}\label{fig:control_affine_approximation}}
    \caption{Bicycle Robot: A slice of the value function for different methods at $t = 2$s. The zero level set (BRT) and the failure set are shown in black and green respectively.}
    \vspace{-2em}
    \label{fig:dubins_brts}
\end{figure}

\begin{figure*}[!t]
    \centering
    \subfloat{\includegraphics[width=0.19\textwidth]{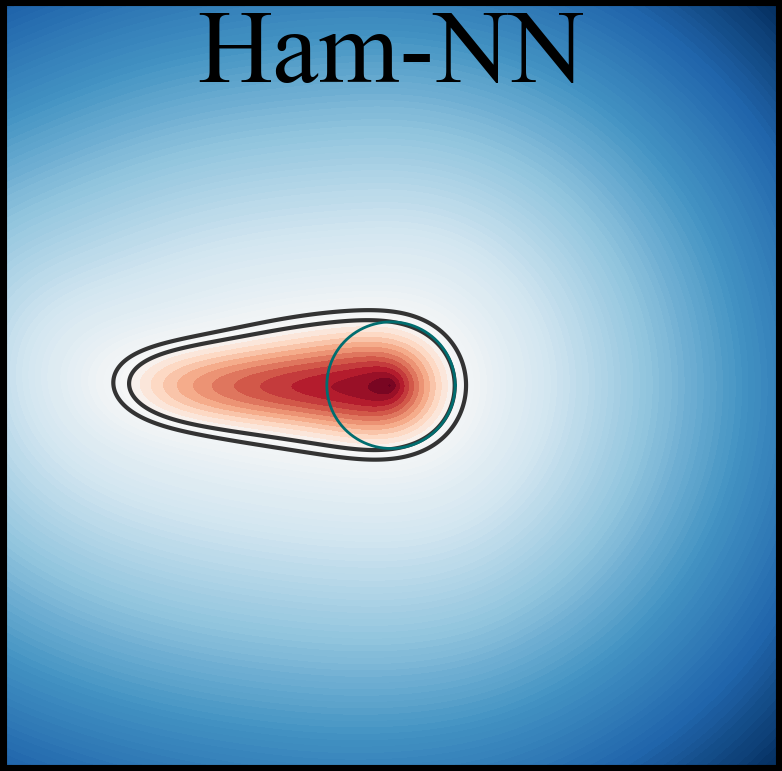}\label{fig:NN_6d}}
    \subfloat{\includegraphics[width=0.19\textwidth]{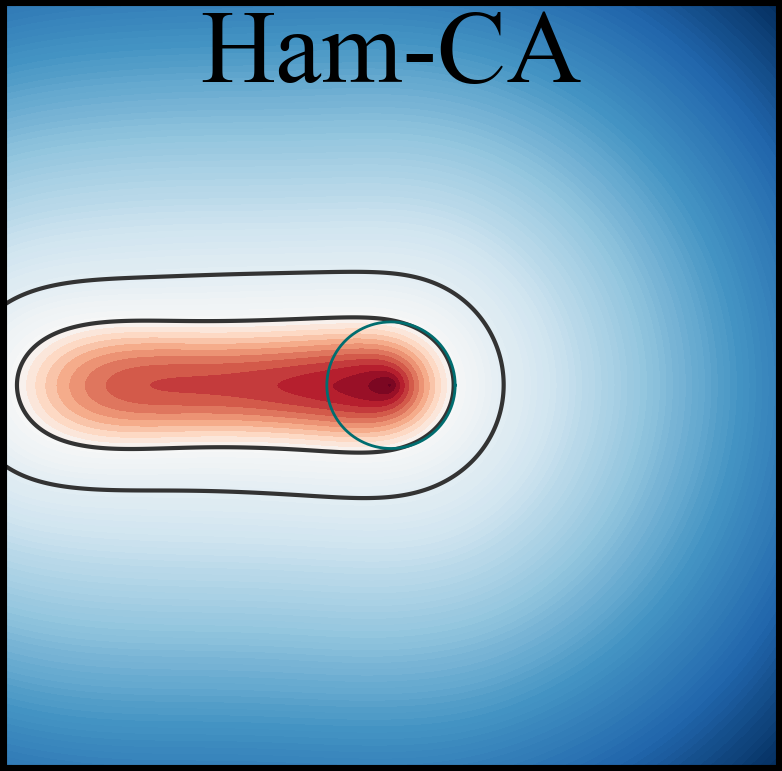}\label{fig:HC_6d}}
    \subfloat{\includegraphics[width=0.19\textwidth]{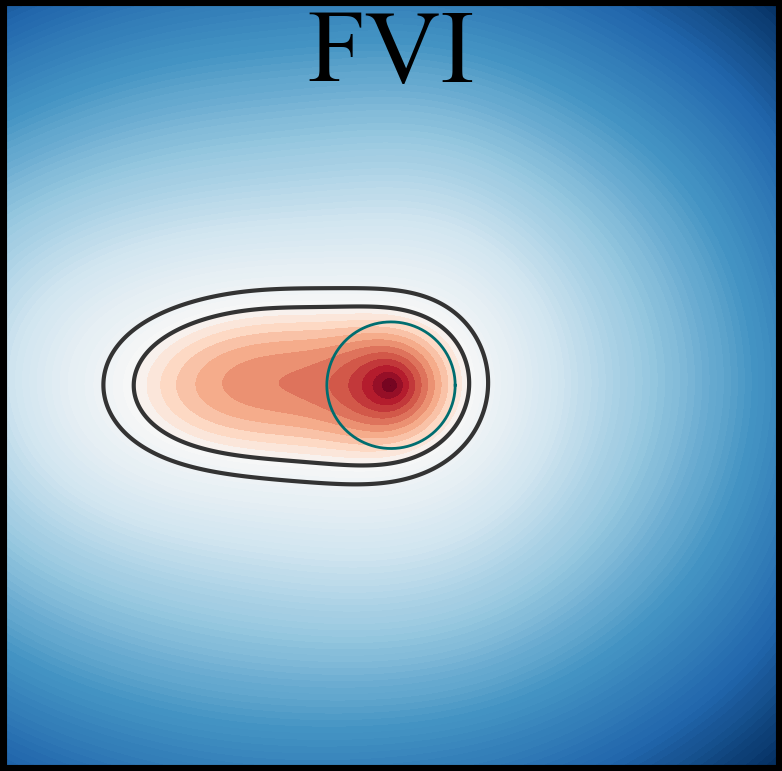}\label{fig:FVI_6d}}
    \subfloat{\includegraphics[width=0.19\textwidth]{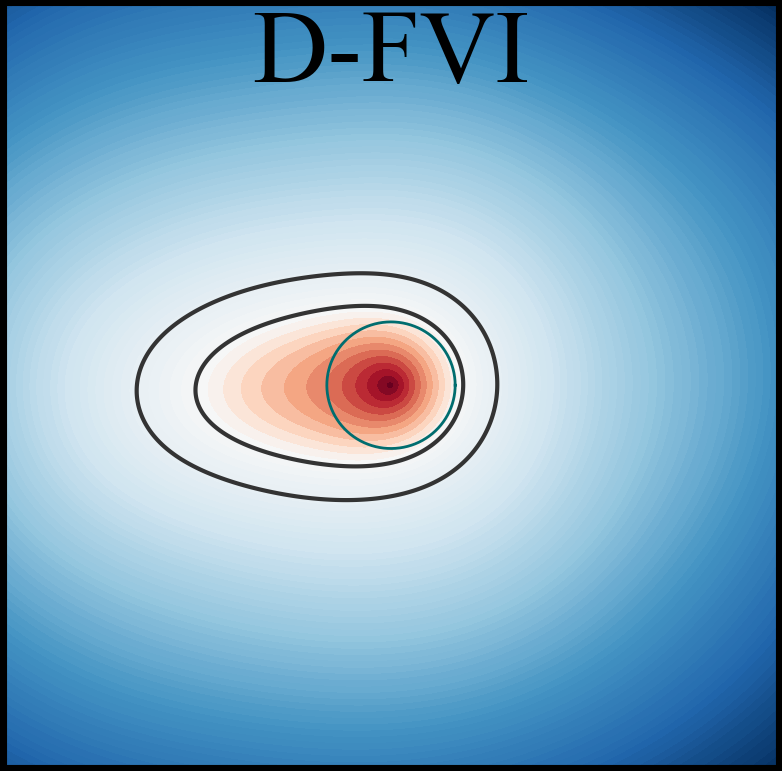}\label{fig:D-FVI_6d}}
    \subfloat{\includegraphics[width=0.19\textwidth]{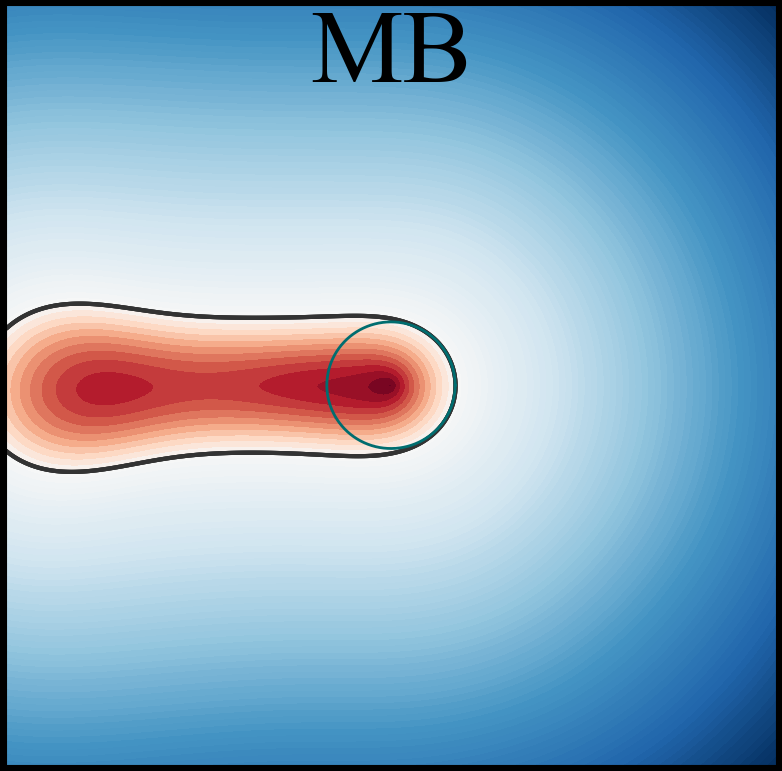}\label{fig:D-mb_6d}}
    \caption{The figures illustrate value function slices (corresponding to initial states $(p_x,p_y,0,12,0,0)$) for the slip-wheel car system. 
    The failure set and sub-$\delta$ level sets (BRTs), with $\epsilon=10^{-2}$ and $10^{-3}$, are also shown. 
    }
    \vspace{-2em}
    \label{fig:comparison}
\end{figure*}

The BRT computation took 45 minutes for the Ham-NN and ground truth methods and around 90 minutes for the Ham-CA method.
This can be explained by the requirement of computing the Hamiltonian at four different control inputs for the Ham-CA method, resulting in an overall higher computation time.
The BRTs obtained using different methods are shown in Fig. \ref{fig:dubins_brts}.
As evident from the figure, both of the proposed variants are able to obtain a high-quality approximation of the value function, without relying on the analytical dynamics of the system.
This is further confirmed by a very low MSE of $4.45 \times 10^{-4}$ for Ham-NN and $1 \times 10^{-12}$ for Ham-CA respectively.
A significantly smaller MSE of Ham-CA can be attributed to the lack of any learning errors that might be present in the Ham-NN method, along with the fact that the dynamics are indeed control-affine in this case, so we expect a recovery of the ground-truth BRT from Ham-CA for small $\Delta$.
This is further aligned with the $0$ FP (\%) for the HAM-CA method (compared to a small but non-zero FP rate of $4.1 \times 10^{-4}$ for the Ham-NN method).

%% file: results.tex
In this section, we conduct a comparative study on two reachability problems: an avoid problem for a 6D slip-wheel vehicle and an avoid problem for a quadruped robot. 
\vspace{0.5em}

\noindent \textbf{\textit{Baselines.}} 
We compare the proposed methods (Ham-NN and Ham-CA) against model-based and model-free baselines. 
Given the high dimensionality of these case studies, we use DeepReach \cite{9561949} to compute the value function and the BRT. 
This will also help illustrate how our method can be combined with different reachability toolboxes.

For model-based comparisons, we use the method from \cite{wang2024providing} that learns an ensemble NN model of the system dynamics along with state-dependent disturbance bound, and then use these dynamics to compute a robust BRT via DeepReach. \textbf{We call this method MB}.

For model-free comparisons, we use a time-dependent fitted value iteration RL method inspired by \cite{fisac2019bridging}. We also compare against a variant of this approach that additionally uses a discount factor $\gamma$ during training.
\textbf{We call these variants FVI and D-FVI, respectively}.
DeepReach-based computations are done using the loss function in (\ref{deepreach_loss}), whereas the RL-based baselines minimize the Bellman error: 
\begin{equation}
\begin{aligned}
    \mathbf{h}&_{\text{FVI}}(x_i,t_i)= \| V\left(x_i, t_i \right)- \\
   &\qquad \qquad \min\left\{l(x_i),\max_{u} V\left(x_{next}^u, t_i+\Delta \right)\right\} \|,\\
    \mathbf{h}&_{\text{D-FVI}}(x_i,t_i)=  \| V(x_i,t_i)-\left[(1-\gamma) l(x_i)+ \right.\\
    &\qquad \quad \left. \gamma \min \left\{l(x_i), \max _{u} V(x_{next}^u,t_i+\Delta )\right\}\right] \|.
\end{aligned}
\end{equation}

\noindent \textbf{\textit{Evaluation Metric.}} Our key evaluation metric is the volume of the verified BRT $\mathcal{B}_\epsilon$, $\mu_{\epsilon}$.
A smaller $\mu_{\epsilon}$ indicates a bigger safe set and hence a better performance.
To quantify this, we sample $N$ distinct states from the state space and count the number of states that fall within $\mathcal{B}_\epsilon$, denoted as $n_{\epsilon}$. The percentage volume of the BRT is given as $\mu_{\epsilon} = 100 \times \frac{n_{\epsilon}}{N}$. We choose a large value of $N = 3 \times 10^6$ to attain samples from a significant portion of the state space.
We use the verification method in \cite{lin2024verification} for all baselines, with $\beta =10^{-10}$ and $\epsilon = 10^{-2}, 10^{-3}$ respectively. 
This corresponds to obtaining a safe set with $99\%$ and $99.9\%$ confidence levels, respectively.

\subsection{Slip-Wheel Car System}\label{section: Slip-Wheel Car System}

In this example, we demonstrate how our methods perform against a black-box non-control-affine system. The dynamics, adapted from \cite{leung2020infusing}, represents a simplified 6D single-track vehicle model with the state $x = [p_x, p_y, \phi, U_x, U_u, r]^T$, where $(p_x, p_y)$ are the Cartesian coordinates, $\phi$ is the yaw angle, $U_x, U_y$ are body frame velocities, and $r$ is the yaw rate.
The control is represented by $u = [\delta, F_x]^T$, where $\delta$ is the steering angle, and $F_x$ is the longitudinal tyre force. 
We further extend the dynamics to account for vehicle sliding when $F_x$ applied exceeds the tire's friction cone. 
Specifically, when $F_{x}^2+F_{y}^2 > \mu F_{z}$, a sliding friction is used to compute the tyre forces. 
Hence, simultaneously applying full brake and steering input will cause sliding, making a bang-bang, extremal controller unfavorable.
The failure set is given by $\mathcal{L} := \left\{x : \sqrt{p_x^2 +p_y^2} \leq 2.5 \right\}$. The time horizon $T$ is 1.5~s.

For all baselines, we use a 3-layer sinusoidal NN with 256 hidden neurons per layer and 100k training iterations. 
The Adam optimizer with a learning rate of $2 \times 10^{-5}$ is employed for training. 
 For Ham-NN, both $H_\nu$ and $u_\psi$ are 2-layer NNs with 128 neurons per layer. 
 A dataset containing 1 million data points is collected to train these models, where $10^{4}$ possible controls are sampled for each point to generate Hamiltonian and optimal control label. 
 The data collection procedure took 20 minutes and training $H_\nu$ took another 15 minutes.
 For FVI and D-FVI baselines, the optimal control is estimated using a $6 \times 6$ grid of possible controls $(\delta, F_x) \in \left[ -\pi/10, \pi/10 \right]  \times \left[ -18794,  5600 \right]$. We set the time step $\Delta=0.002s$ and apply a discount factor annealing scheme where $\gamma$ starts at $0.99$ and gradually increases to $0.999$ during training. 

\begin{table}[h!]
\centering
\vspace{-0.5em}
\begin{tabular}{|l|lp{1.05cm}p{0.7cm}p{0.8cm}p{0.65cm}|}
                          \hline       & Ham-NN     & Ham-CA      & FVI    & D-FVI & MB     \\ \hline
$\mu^{car}_{\epsilon = 10^{-2}}$ &  \textbf{3.48\%} & 3.91\% & 5.03\% & 3.78\%  & 5.56\% \\
$\mu^{car}_{\epsilon = 10^{-3}}$ & \textbf{4.33\%} & 10.30\% & 7.43\% & 8.50\%  & 5.56\% \\
Training time                       & 1.5h total &   \textbf{0.5}h   & 7.8h &  9.5h    &   2.0h  \\ \hline
$\mu^{quadruped}_{\epsilon = 10^{-2}}$ &  \textbf{6.87\%} & 16.96\% & 14.51\% & 12.27\%  & 11.03\% \\
$\mu^{quadruped}_{\epsilon = 10^{-3}}$ & \textbf{8.36\%} & 34.01\% & 34.29\% & 30.82\%  & 17.50\% \\
Training time            & 2.5h total &   \textbf{2h}   & 7h &  7h    &  3h  \\
\hline
\end{tabular}
\caption{The BRT volume ($\mu$) and training time for the slip-wheel car (top) and the quadruped (bottom) case studies.}
\vspace{-0.5em}
\label{table:verification results}
\end{table}
The verified volume of the BRT under different methods, as well as total training time, is shown in Table~\ref{table:verification results}.
Ham-NN consistently outperforms other methods in achieving a lower BRT volume (a higher safe set volume).
Ham-CA requires the least training time as it bypasses the need for training a $H_\nu$. In addition, it synthesizes a conservative BRT because the Hamiltonian estimated via hypercube sampling is always an underapproximation of the actual one. 
In spite of the conservatism, the volume of verified BRT still expands as the required safety level increases (i.e., $\epsilon$ decrease), which is illustrated by the growing unsafe set contours in Fig.~\ref{fig:HC_6d}.
This suggests that Ham-CA is more sensitive to learning errors.
Another variant giving a conservative BRT is the model-based method since it captures the uncertainty in the dynamics model. This uncertainty, while allows it to safeguard against potential modeling errors, can lead to overly conservative behavior.
This is also evident from the same BRT volume of MB for the two $\epsilon$ levels -- while MB is over-conservative when the safety criterion is loose, it yields the second-smallest unsafe set when $\epsilon = 10^{-3}$ since the conservatism makes it more resilient to learning errors.
On the other hand, the D-FVI variant achieves a similar unsafe volume compared to the Ham-NN at $\epsilon = 10^{-2}$, as it samples a grid of controls to estimate the optimal one. However, this advantage comes at the cost of a much longer training time.
Moreover, as we increase the required safety assurance level, the verified BRT volume increases significantly, indicating higher learning errors in D-FVI. 
In contrast, the BRT volume increases only modestly for Ham-NN, highlighting the robustness of the proposed approach.
\begin{figure*}[!t]
    \centering
    \subfloat[]{\includegraphics[width=0.3916\textwidth]{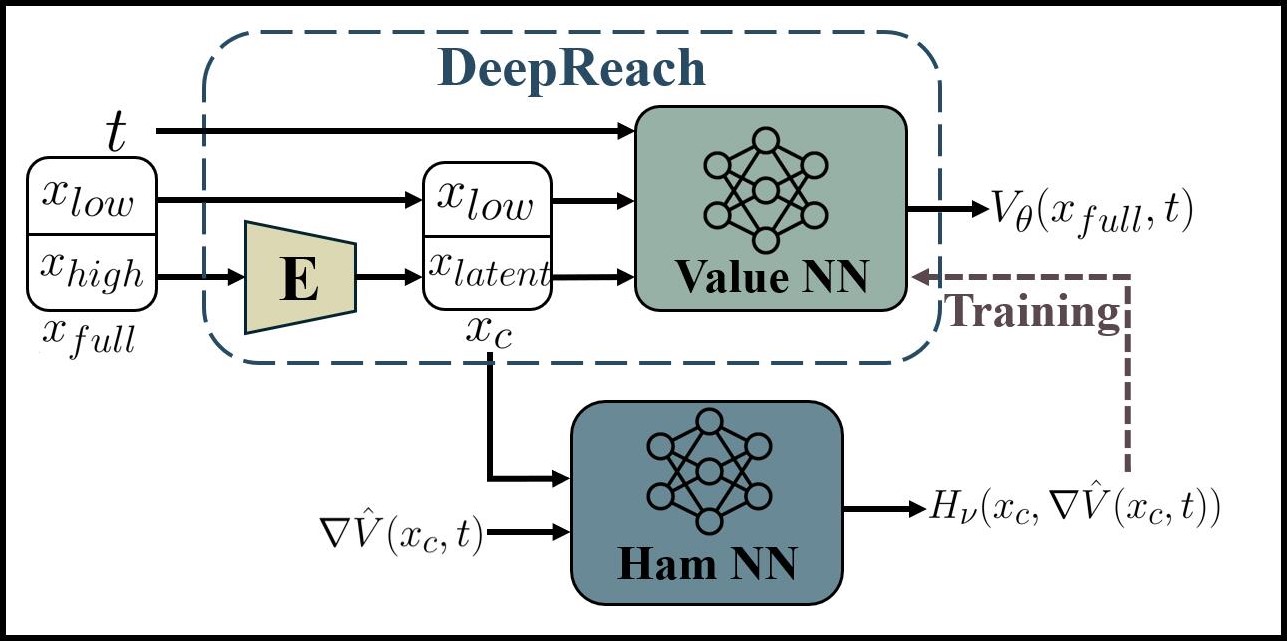}\label{fig:architecture}}
    \subfloat[]{\includegraphics[width=0.1949\textwidth]{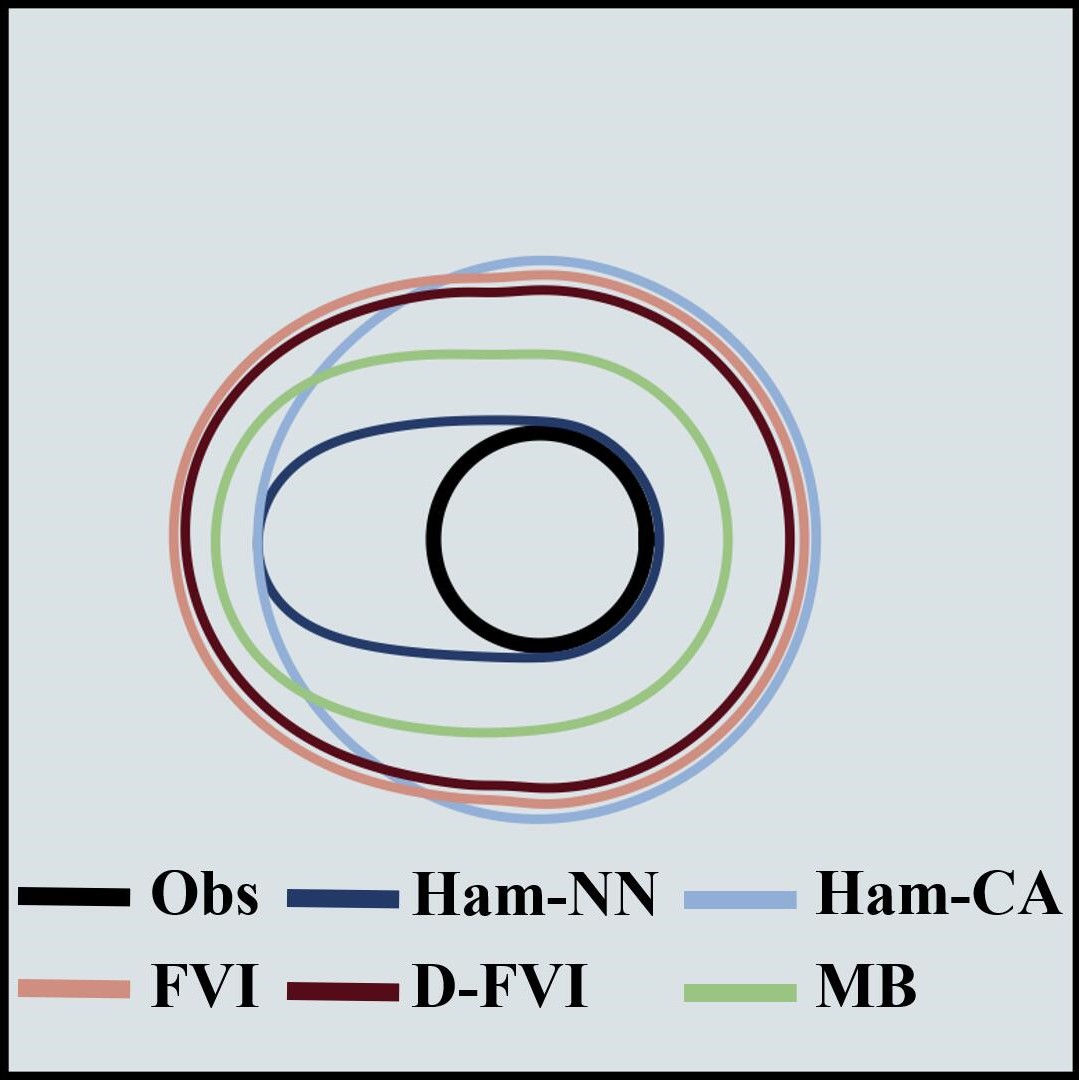}\label{fig:overlay_brts}}
    \subfloat[]{\includegraphics[width=0.1987\textwidth]{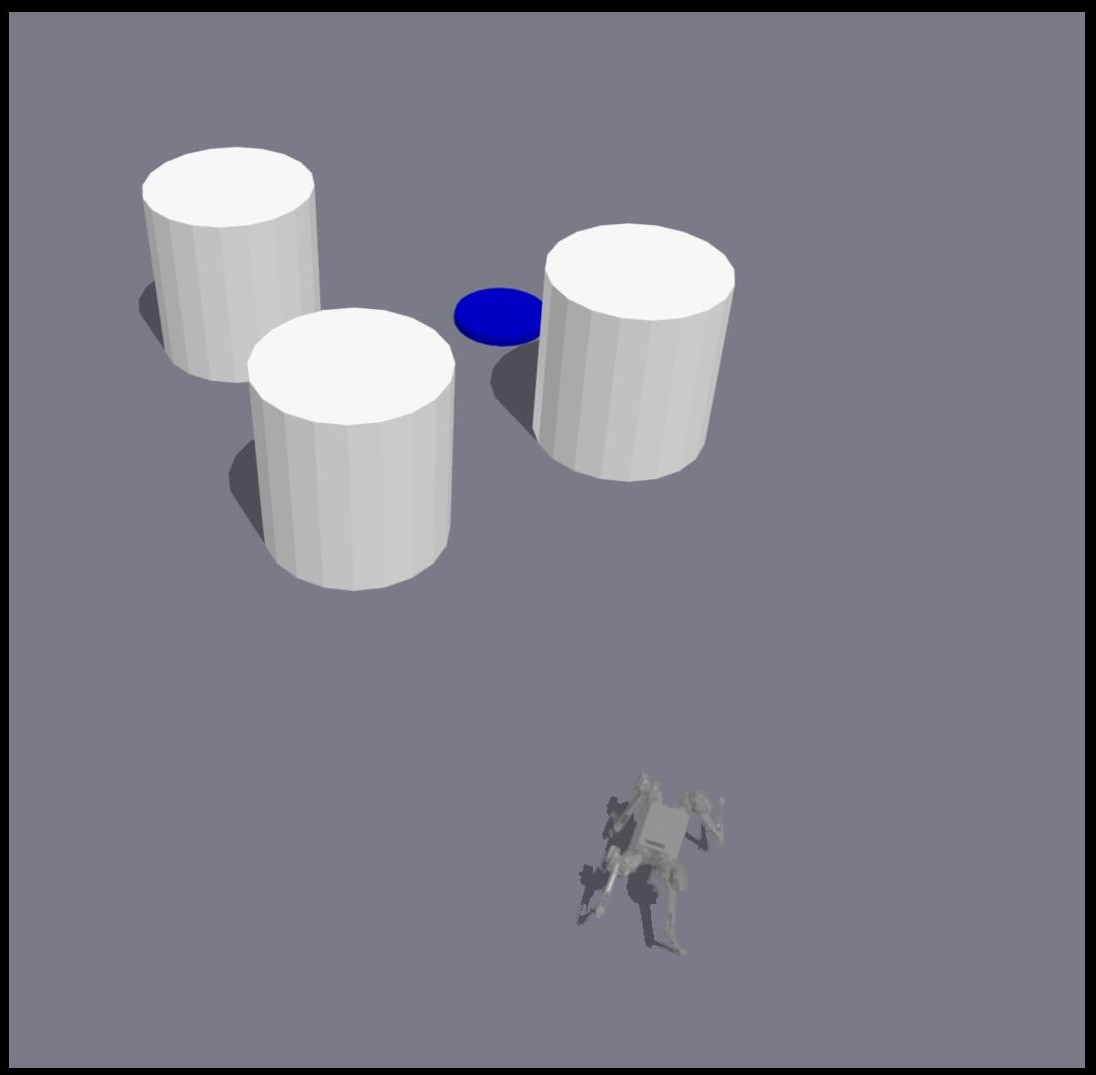}\label{fig:sim_env}}
    \subfloat[]{\includegraphics[width=0.195\textwidth]{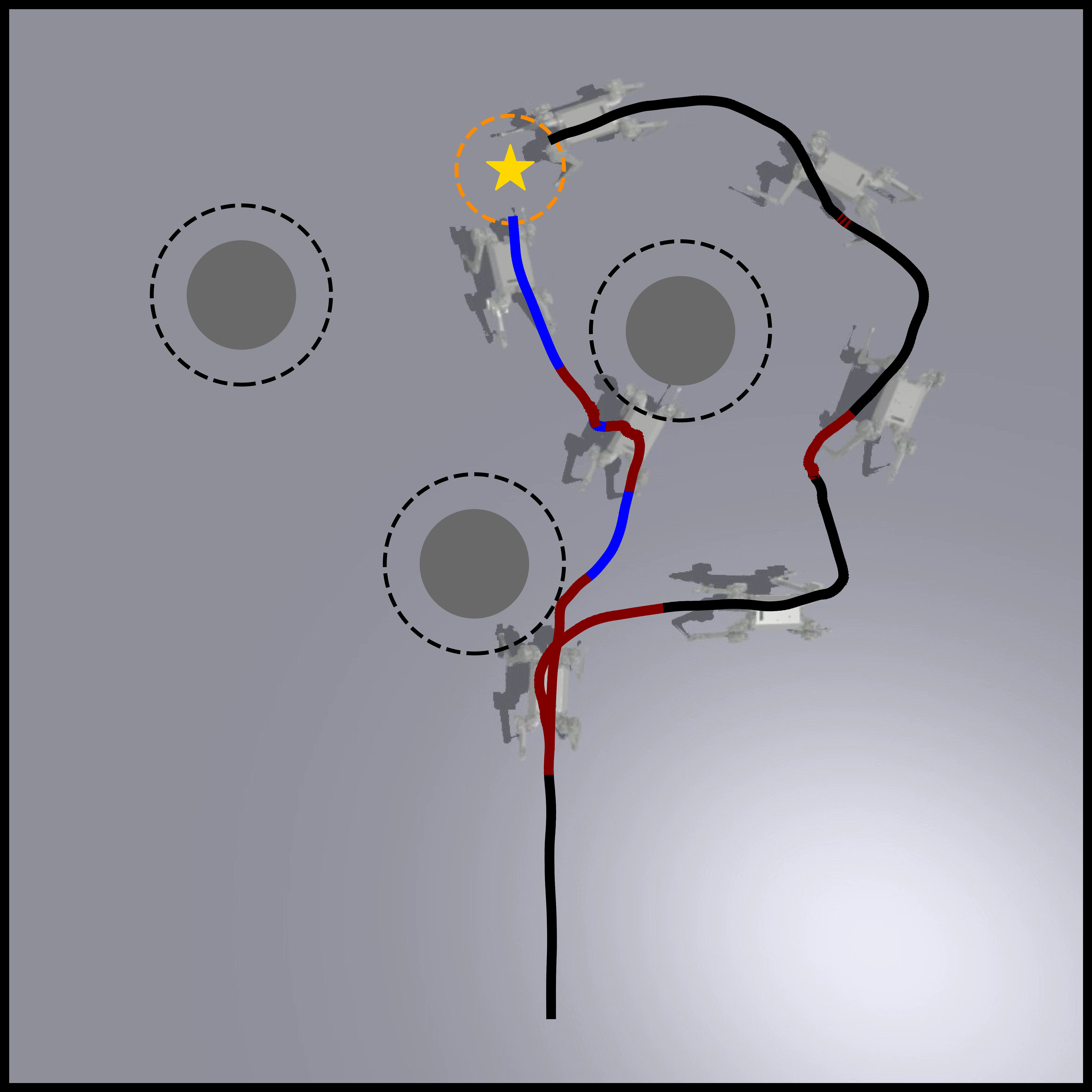}\label{fig:traj}}
    \caption{(a) An overview of value function prediction: DeepReach takes in the full state $x_{full}$, internally compresses it using an encoder, and feeds the condensed state $x_c$ to its safety value NN. The Hamiltonian estimation NN also takes $x_c$ as input. (b) The BRTs verified with $\epsilon = 10^{-3}$ for different variants. Ham-NN consistently outperforms all other baselines. (c) A snapshot of the Issac Gym environment that is used to learn the BRT and for deploying the safety filter. (d) The safety-filtered trajectories for the quadruped example. The obstacle contours are shown in black dashed lines, and the star represents the goal position. Maroon line segments indicate the activation of the safety controller. Due to the over-conservatism of the MB approach, the robot takes an inefficient detour.
    }
    \vspace{-2em}
    \label{fig:comparison}
\end{figure*}
\subsection{Quadruped System}
We take the last example to demonstrate how the proposed method can be applied to complex, high-dimensional systems. We use Isaac Gym \cite{makoviychuk2021isaac} to simulate the robot and the underlying dynamics are unavailable. 
We leverage a pre-trained Reinforcement Learning (RL) policy from \cite{margolis2024rapid} for low-level control. 
The RL policy takes a high-level twist commands $u_h=[v_x^c, \omega^c]$ along with the robot's state $x_{full}$ as inputs and outputs the desired joint angles to track $u_h$. 

To simplify the problem, the state $x_{full}$ is divided into low-frequency, important states $x_{low}$ and high-frequency, less critical ones $x_{high}$. 
Here, $x_{low}=(p_x, p_y, \phi, v_x, v_y, \dot{\phi} )$ includes robot's COM position, yaw angle, body frame velocities, and yaw rate, while $x_{high}=\left[ \boldsymbol{q}, \dot{\boldsymbol{q}}, \boldsymbol{g}_p, \boldsymbol{c}, u_h^{last} \right]$ include the joint positions and velocities, projected gravity, foot contacts, and the previous high-level command.
Since the RL controller follows specific gaits to track the twist command, we train an autoencoder to compress the $x_{high}$ into a 2-dimensional latent representation $x_{latent} = \left[z_1,z_2\right]$. 
The low-frequency state and the latent state are concatenated to create a condensed state representation $x_c=[x_{low}, x_{latent}]$ for the quadruped.
Here, DeepReach can be thought of as a composition of the encoder and the safety value NN $V_\theta$, with $x_{full}$ being its input.
A detailed illustration is provided in Fig.~\ref{fig:architecture}.
The failure set is defined as $\left\{x_c : \sqrt{p_x^2 +p_y^2} \leq 0.5 \right\}$, which corresponds to an obstacle of radius 0.5.  

All baselines use a 3-layer sinusoidal NN with 512 hidden neurons per layer, while other hyperparameters remain the same as in Section~\ref{section: Slip-Wheel Car System}. 
We collect a dataset of 1 million data points to train $H_\nu$ and $u_\psi$, which are 2-layer MLPs with 128 neurons per layer. 
Due to the relatively slower simulation queries in this case, we only sample 6 possible $u_h$, including the vertices of the control hypercube, to determine the optimal control.

The obtained results are presented in Table~\ref{table:verification results}. Similar to the slip-wheel system, Ham-NN outperforms all other baselines across both $\epsilon$ values. 
Given the higher complexity of the quadruped system, we see an even bigger gap between the performance of Ham-NN and other methods.
Notably, both Ham-NN and MB exhibit better robustness to changes in $\epsilon$ compared to other variants. 
The MB variant is more robust due to its conservatism. 
For Ham-NN, we attribute this to the smoothening effect of the Hamiltonian estimator $H_\nu$ -- the NN effectively smoothens out the sudden changes in the Hamiltonian in the parts of the state space where the dynamics are stiff, resulting in a simplified reachability problem for DeepReach and lower learning errors. 
Conversely, the Ham-CA variant is more sensitive to sudden changes in Hamiltonian and, therefore, performs worse than Ham-NN. 

Lastly, we demonstrate the utility of the verified BRT by using it as a safety filter for a simple PD controller designed to reach a goal position in a cluttered environment (see Fig.~\ref{fig:traj}). 
We apply a least-restrictive safety filter where the safety control is activated if the value function predicted at the current state falls below a specified threshold; otherwise, the nominal controller is applied \cite{borquez2024safety}.
This threshold is set to the $\delta$ value corresponding to $\epsilon = 10^{-3}$. 
A comparison of the trajectories resulting from the learned solution using the proposed method and the model-based approach is shown in Fig.~\ref{fig:traj}. 
We do not show other methods because their BRT is too conservative and takes the robot out of the operating domain as a result.
The filtered trajectory with our method successfully identifies a gap to traverse through, while the model-based method is overly conservative and detours to reach the goal, resulting in sub-optimal performance.

%% file: conclusion.tex
We propose a framework to compute BRTs and safety controllers for black-box dynamical systems. Our approach can be used with existing reachability toolboxes to solve the HJB-VI for any general black-box system.
We further propose a time-efficient variant (Ham-CA), which gives accurate solutions for control-affine systems and conservative BRT approximations for general dynamics.
Through various case studies, we demonstrate the effectiveness of the proposed approach in recovering safe regions for black-box systems.

While the current work can be seamlessly integrated with learning-based level-set methods, its capability to solve HJB-VI is inherently constrained by the former. Therefore, developing more sophisticated and robust learning techniques would facilitate the proposed method to encompass higher-dimensional and more intricate reachability problems for black-box systems. In addition, the input size of Ham-NN, being double the system dimension, poses substantial challenges when dealing with extremely high-dimensional problems. Mitigating this limitation is another crucial direction for future research.